\documentclass[aps,amssymb,pra,superscriptaddress,showpacs,preprint]{revtex4}
\usepackage{epsfig}
\usepackage{bm}
\usepackage{ulem}

\begin{document}

\title{Hysteresis effects in rotating Bose-Einstein condensates}
\author{B.~Jackson}
\affiliation{Dipartimento di Fisica, Universit\`a di Trento and CNR-INFM R\&D
 Center on Bose-Einstein Condensation, I-38050 Povo, Italy}
\affiliation{School of Mathematics and Statistics, University of 
 Newcastle upon Tyne, NE1 7RU, United Kingdom}
\author{C.~F.~Barenghi}
\affiliation{School of Mathematics and Statistics, University of 
 Newcastle upon Tyne, NE1 7RU, United Kingdom}
          
\begin{abstract}
 We study the formation of vortices in a dilute Bose-Einstein condensate 
 confined in a rotating anisotropic trap. We find that the number of vortices 
 and angular momentum attained by the condensate depends upon the rotation 
 history of the trap and on the number of vortices present in the 
 condensate initially. A simplified model based on hydrodynamic equations is 
 developed, and used to explain this effect in terms of a shift in the 
 resonance frequency of the quadrupole mode of the condensate in the 
 presence of a vortex lattice. Differences
 between the spin-up and spin-down response of the condensate are 
 found, demonstrating hysteresis phenomena in this system.
\end{abstract}
\date{\today}


\maketitle

\section{Introduction}

The realization of Bose-Einstein condensation (BEC) in ultracold 
gases has provided a powerful new system in the study of superfluids, 
combining better experimental control and theoretical tractability than 
liquid helium. A particularly
striking example concerns the response of the fluid to rotation. 
Due to the irrotationality property of superfluids, application of a rotating 
perturbation results in the creation of a lattice of quantized vortex lines,
a process that was originally observed in He-II \cite{donnelly}. 
Analogous experiments in ultracold
gases have found the formation of similar structures in Bose condensates 
\cite{madison00,chevy00,abo-shaeer01,raman01,hodby02,haljan01,madison01}, and 
recently in a fermionic gas that is superfluid due to Cooper pairing 
between atoms \cite{zwierlein05}. 

An interesting question raised by these experiments concerns the angular
velocity of the rotating potential required to nucleate vortices. Since 
the energy per particle of a condensate containing a vortex, $E_v$, exceeds 
that of the ground
state, $E_0$, by transforming to a frame rotating with angular velocity 
$\Omega$ it is straightforward to show that vortices are energetically 
favored if $\Omega>(E_v-E_0)/\hbar$ \cite{dalfovo96}. However, higher rotation
rates are generally required to nucleate vortices in experiments
\cite{madison00,raman01,hodby02,madison01}, which reflects the existence of
an energy barrier to vortex formation \cite{fetter01}. This can be overcome 
by excitation of collective modes localized at the surface of the condensate 
\cite{dalfovo00,kraemer02}.

The presence of two different critical angular velocities implies the 
existence of metastable states, where vortices are energetically favorable 
but dynamically are unable to form, and in turn the possibility of hysteresis 
phenomena \cite{dalfovo00,garciaripoll01a}.
Hysteresis has long been observed in rotating superfluid helium. For example, 
experiments have found a significant difference between the number of vortices
obtained when the fluid is 
spun up to when it is spun down \cite{packard72}. Reproducible hysteresis loops
were measured by minimizing the mechanical and thermal noise of the apparatus 
\cite{mathieu80}. Jones {\it et al.} \cite{jones95} attempted to explain this 
effect by using a principle of local momentum conservation. Fortunately atomic
Bose-Einstein condensates are less noisily coupled to their environment than 
rotating liquid helium, and, unlike liquid helium, can be studied using a 
model that has great predictive power, namely the Gross-Pitaevskii (GP) 
equation. 
 
In this article we discuss a particular example of hysteresis, where the 
amount of circulation already present in a Bose condensate influences the 
final angular momentum and number of vortices that can be attained. This 
effect has been exploited in the experiment of Ref.\ \cite{bretin04} to rotate
the condensate to high angular velocities. The aim of this paper is to
explore this issue theoretically
in more detail. Following 
previous studies \cite{feder01,lundh03,kasamatsu03,lobo04,parker05a,parker06} 
we solve the GP equation for the condensate wavefunction $\Psi({\bm r},t)$ 
to study the vortex formation process in a rotating anisotropic trap. 
However, we go further by considering cases where the trap rotation
frequency is changed during the simulation, as well as when vortices are 
already present in the condensate. Moreover, new insight is gained by
comparing our findings to the results of a hydrodynamic model 
\cite{cozzini03} of the condensate, which yields a 
resonance in the angular momentum transfer associated with excitation 
of the quadrupole collective mode. The presence of a vortex lattice then has
the effect of shifting this mode frequency and therefore the resonance
to higher values. As 
a further example of hysteresis we consider the case where the trap rotation
frequency is slowly ramped up and then down, demonstrating a difference
in the trajectories of the angular momentum between the two cases. 
 
We note that the discussion in this paper is restricted to very low 
temperatures, where the thermal cloud is insignificant and the condensate
dynamics can be accurately modelled with the GP equation.
An interesting issue concerns the effect of finite temperatures on the
phenomena described here. However, to address this question requires one to
treat the dynamics of both the condensate and thermal cloud consistently,
since the thermal cloud also responds to a rotating potential by spinning up 
during a timescale related to the frequency of collisions between the atoms 
\cite{gueryodelin00}. Treatment of the coupled dynamics is conceptually rich 
and numerically intensive, and is outside the scope of this work.

\section{Vortex formation}

The GP equation is solved numerically for a harmonically 
trapped condensate in 2D, which corresponds to the case where the axial trap 
frequency is much larger than the mean field interaction energy, such that
motion along the axial direction is frozen. 
Similar 2D studies \cite{lundh03,parker06} have previously found good 
qualitative, and in some cases quantitative,
agreement with experimental data even when this condition is not 
satisfied \cite{madison00,hodby02}, confirming that 
much of the crucial physics of the vortex formation process is captured in 2D.

Angular momentum is imparted to the condensate by rotating 
an elliptically-deformed harmonic trap, which is represented by the 
potential
\begin{equation}
 V ({\bm r}) = \frac{1}{2} m\omega_{\perp}^2 \left[ (1+\epsilon) x'^2 +
 (1-\epsilon) y'^2 \right] ,
\label{eq:trap}
\end{equation} 
where a rotational transformation of the $(x,y,z)$ Cartesian coordinate system
is used, such that  
$x'=x\cos(\Omega t)+y\sin(\Omega t)$ and $y'=-x\sin(\Omega t)+y\cos 
(\Omega t)$ correspond to rotation of the trap at frequency $\Omega$.
Dimensionless units are also used, with the units of length, time and 
energy given by $(\hbar/(m\omega_{\perp}))^{1/2}$, $\omega_{\perp}^{-1}$ and
$\hbar \omega_{\perp}$ respectively. The GP equation then becomes
\begin{equation}
 i \frac{\partial}{\partial t}  \Psi = \left\{ \frac{1}{2} \left [-\nabla^2 +
 (1+\epsilon)x'^2+(1-\epsilon)y'^2 \right ] + 
 g |\Psi|^2 \right\} \Psi .
\label{eq:GP-eqn}
\end{equation}
Mean field interactions are represented by $g=4\pi N' a$, where $N'$ is the 
number of atoms per unit length in the axial $z$ direction, while $a$ denotes 
the $s$-wave 
scattering length. Throughout this paper we will use $g=450$, although the 
results are expected to be generally applicable to other interaction 
strengths.

To provide the initial condition for the simulation, Eq.~(\ref{eq:GP-eqn}) 
is numerically propagated in imaginary time with $\epsilon=0$ and $\Omega=0$,
such that the wavefunction converges to the condensate ground state without 
vortices. In order to model vortex formation the simulation is then run in 
real time with a rotating elliptical trap ($\epsilon=0.1$, $\Omega>0$), where 
the deformation $\epsilon$ is switched on 
abruptly at $t=0$, rather than being turned on gradually. 

\begin{figure}[h]
\centering \scalebox{0.45}
 {\includegraphics{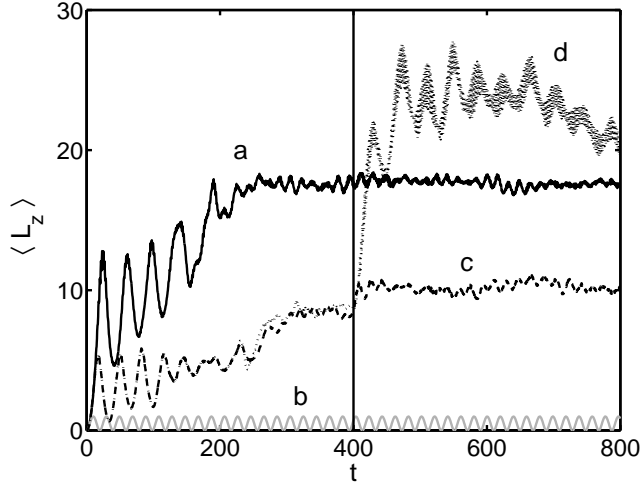}}
 \caption{Angular velocity, $\langle L_z \rangle$ (in units of $\hbar$) as a 
 function of time, $t$ (units of $\omega_{\perp}^{-1}$) for $\epsilon=0.1$ 
 and no vortices present in the condensate initially. 
 The solid lines plot the results when the initial trap 
 rotation frequency $\Omega$ (in units of $\omega_{\perp}$) is maintained up 
 to $t=800$, with $\Omega=0.78$ 
 (curve a, black) and $\Omega=0.9$ (curve b, grey). The broken lines (curves 
 c and d) represent cases where $\Omega=0.7$ up to $t=400$ (marked by a 
 vertical line), after which 
 the rotation frequency is changed to $\Omega=0.78$ (c, dashed) or 
 $\Omega=0.9$ (d, dotted). Note that all of the figures in this 
 paper are plotted in dimensionless units.}
\label{fig:ang-rot2}
\end{figure}

Fig.~\ref{fig:ang-rot2} illustrates the subsequent time-dependent response of 
the condensate by plotting the mean angular momentum 
$\langle L_z \rangle = \int d \bm{r}\, \Psi^* \hat{L}_z \Psi$,  
with $\hat{L}_z=i[y \partial_x - x \partial_y]$. If $\Omega=0.78$ (curve a)
one sees that the angular momentum increases initially, undergoing large 
oscillations before settling to an almost constant value for $t>250$, similar 
to behavior found in previous studies \cite{lundh03,parker05a,parker06} as 
well as experiments \cite{madison01}. The 
initial oscillations correspond to variations in the quadrupolar 
deformation of the condensate which subsequently diminish as vortices enter 
the condensate. The vortices then continue to
undergo complicated dynamics, albeit in such a way that the angular momentum 
remains in quasi-equilibrium. However, the vortices do not crystallize to 
form a stable, ordered lattice, which would require the inclusion of 
dissipation 
\cite{kasamatsu03,lobo04,tsubota02,penckwitt02} or integration of the GP 
equation over much longer timescales \cite{parker05a,parker06}.   

If the trap rotation rate is significantly smaller or larger than
$\Omega=0.78$ then the condensate angular 
momentum displays a periodic time dependence (see Fig.~\ref{fig:ang-rot2} 
curve b, for 
$\Omega=0.9$). This behavior  
corresponds to an oscillation of the quadrupolar deformation of the condensate,
$\delta= \langle y'^2-x'^2 \rangle / \langle y'^2+x'^2 \rangle$. The 
angular momentum is found to be close to the value 
$\langle L_z \rangle = \Omega \Theta$ if the momentum of 
inertia is given by the expression for a irrotational superfluid, 
$\Theta=\delta^2 \langle x'^2+y'^2 \rangle$ \cite{zambelli01,garciaripoll01b}.
Hence the 
condensate remains irrotational and no vortices are nucleated, in contrast to
the case where $\Omega=0.78$. Note that the oscillations in the quadrupolar 
deformation arise due to the sudden imposition of the rotating trap's 
deformation, and their amplitude is lower if $\epsilon$ is turned on 
gradually.    

In the two examples given, the trap rotation rate was maintained at a constant
value throughout the simulation. If, instead, the rotation rate is changed to 
another value 
midway through the run, then very different results are found. Curves c and
d of Fig.~\ref{fig:ang-rot2} represent simulations where for 
$0<t<400$ the 
trap rotates at $\Omega=0.7$, resulting in vortex nucleation. At $t=400$ the 
rotation rate is abruptly changed to either $\Omega=0.78$ (curve c) or 
$\Omega=0.9$ (curve d). In 
the former case the angular momentum increases only slightly, and remains much
smaller ($\langle L_z \rangle \simeq 10$) than in the 
case where $\Omega=0.78$ from the beginning ($\langle L_z \rangle \simeq 18$). 
For $\Omega=0.9$ (curve d), however, the angular 
momentum jumps to a much larger value (corresponding to a major increase in
the number of vortices), which is in stark contrast with the lack of vortices 
when the condensate is rotated at this frequency initially (curve b).
These examples demonstrate that the vortex formation process strongly depends 
upon the ``rotation history'' of the condensate.  
 
To explore this question further, we have also conducted simulations where the
trap rotation frequency remains constant throughout each run, but the initial 
state already contains vortices. These states can be obtained by solving 
the GP equation for a non-inertial frame of reference which
rotates at constant angular velocity $\Omega_0$, where $t$ is now 
imaginary
\begin{equation}
 i \frac{\partial}{\partial t}  \Psi = \left[ \frac{1}{2} \left (-\nabla^2 +
   x^2 + y^2 \right) + g |\Psi|^2  -
   \Omega_0 \hat{L}_z \right] \Psi .
\label{eq:GP-rot}
\end{equation}
Hereafter we will use $\Omega_0$ to distinguish the rotation of the reference
frame (in which we solve the GP equation in imaginary time) from the (real 
time) trap rotation rate $\Omega$. For 
sufficiently large $\Omega_0$, imaginary time propagation leads to the 
appearance of vortices, which eventually settle into an ordered lattice to 
yield the stationary solution for this rotating frame.

\begin{figure}[h]
\centering \scalebox{0.45}
 {\includegraphics{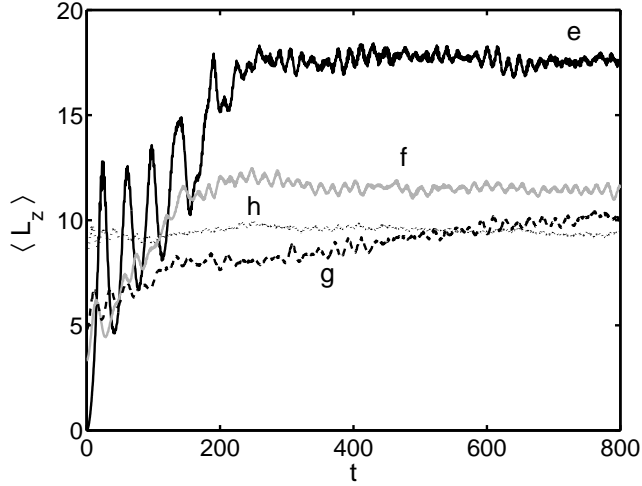}}
 \caption{Angular momentum as a function of time
 for a condensate in a rotating trap with
 $\epsilon=0.1$ and $\Omega=0.78$. The various simulations use
 different initial conditions, which correspond to stationary states for an 
 isotropic condensate in a frame rotating with angular velocity $\Omega_0$. 
 Plotted are $\Omega_0=0$ (curve e, solid black line), $\Omega_0=0.5$ 
 (curve f, grey), 
 $\Omega_0=0.6$ (curve g, dashed) and $\Omega_0=0.78$ (curve h, dotted).} 
\label{fig:ang-rot}
\end{figure}

The wavefunction thus found can be used as the initial condition for
Eq.~(\ref{eq:GP-eqn}), which is integrated in real time as in the previous 
section. Fig.~\ref{fig:ang-rot} shows the 
resulting time evolution of the angular momentum with $\Omega=0.78$ for 
different initial states. The black solid line (curve e) corresponds to the 
same run represented by curve a in Fig.~\ref{fig:ang-rot2}, where the 
condensate contains no vortices initially ($\Omega_0=0$). The grey line
(curve f), by way of contrast, is for $\Omega_0=0.5$, where the initial state 
contains six vortices and possesses an angular momentum of $\langle L_z \rangle
= 3.64$. The subsequent rotation of the trap at $\Omega=0.78$ leads to
a final angular momentum of $\langle L_z \rangle \simeq 11$, which is only 
around 60~\% of the value attained when
no vortices are present initially. If higher values of $\Omega_0$ 
(corresponding to more initial vortices) are used, this 
deficit in the angular momentum becomes even larger. Indeed, if 
$\Omega_0=0.78$, the angular momentum remains almost 
constant throughout the evolution in real time. 

\begin{figure}[h]
\centering \scalebox{0.75}
 {\includegraphics{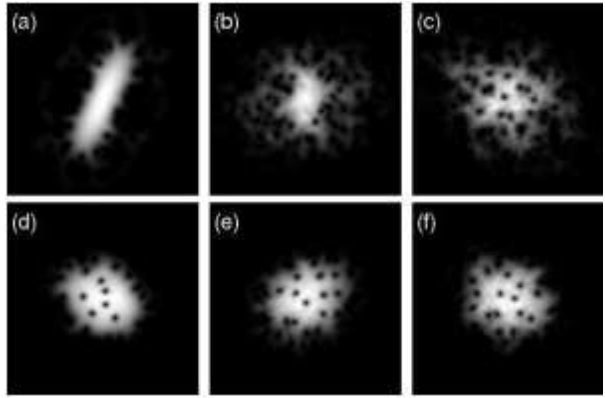}}
 \caption{Condensate density $|\Psi(\bm{r},t)|^2$ with a trap rotating at
 angular 
 velocity $\Omega=0.78$ for (a) $t=100$, (b) $t=200$, and (c) $t=400$, with 
 $\Omega_0=0$. Panels (d)-(f) show a similar simulation but with 
 $\Omega_0=0.6$, for (d) $t=50$, (e) $t=200$, (f) $t=400$. The holes in 
 the density profiles correspond to vortices.}
\label{fig:panels3}
\end{figure}

These differences in angular momentum reflect similar variations in the 
number of vortices present in the condensate. This is illustrated in 
Fig.~\ref{fig:panels3}, where (a)-(c) show snapshots of the condensate density,
$|\Psi({\bm r}, t)|^2$, at various times for $\Omega_0=0$ and $\Omega=0.78$.
The corresponding case for $\Omega_0=0.6$ is shown in Figs.~\ref{fig:panels3} 
(d)-(f); vortices are already present initially, but the 
final number of vortices at $t=400$ is less than for $\Omega_0=0$ (comparing
Figs.~\ref{fig:panels3} (c) and (f)).
       
\begin{figure}[h]
\centering \scalebox{0.45}
 {\includegraphics{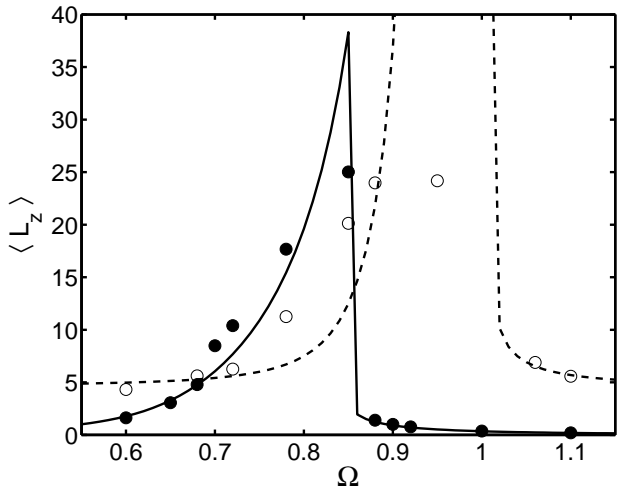}}
 \caption{Maximum angular momentum attained by a condensate in an elliptical
 trap rotating with frequency $\Omega$ ($\epsilon=0.1$). The solid and open 
 circles show the results for $\Omega_0 = 0$ and $\Omega_0 =0.5$ respectively.
 The lines plot the maximum angular momentum found by solving the hydrodynamic
 model (\ref{eq:en1}-\ref{eq:en8}) for $\Omega_0=0$ (solid) and 
 $\Omega_0=0.5$ (dashed).} 
\label{fig:angmax}
\end{figure}

To help explain these differences, in Fig.~\ref{fig:angmax} we plot the 
angular momentum as a function of the trap rotation frequency for 
$\Omega_0=0$ and $\Omega_0=0.5$. If there are no vortices and 
$\langle L_z \rangle$ has a periodic time dependence
then the plotted value is the peak value between $t=0$ 
and $t=400$. If vortices are present the value of $\langle L_z \rangle$ 
is taken at $t=400$, at which point the angular momentum has generally 
plateaued at an approximately constant value. The filled circles represent
the results for the initial condition $\Omega_0=0$, and it is apparent that 
appreciable vortex formation
(and hence angular momentum transfer) occurs in the approximate range 
$0.68 < \Omega <  0.88$. Furthermore, within this range the angular momentum 
rises with increasing $\Omega$, in agreement with previous experimental
\cite{hodby02} and theoretical \cite{parker06} studies,
where a similar dependence was found when plotting the final number of 
vortices against $\Omega$.  

The angular momenta attained at $t=400$ when the initial condition
is $\Omega_0=0.5$ are plotted in Fig.~\ref{fig:angmax} with open circles
\cite{footnote1}. We see that, compared to the $\Omega_0=0$ case,
appreciable increases in 
$\langle L_z \rangle$ tend to take place at higher $\Omega$ .
Hence, nucleation of additional vortices
occurs at higher rotation frequencies when vortices are already present in the 
condensate. This effect accounts for the differences in angular momentum
apparent in Figs.~\ref{fig:ang-rot2} and \ref{fig:ang-rot}. 

\section{Hydrodynamic model}

The numerical solutions of the Gross-Pitaevskii equation discussed in the 
previous section accurately treat the dynamics of the condensate at very low
temperatures. However, it is also instructive to consider a simplified, 
approximate model that is much less numerically intensive to solve than the 
GP equation. This allows us to more easily explore parameter space, as well as 
providing new insight into the phenomena observed so far.
Our simplified model is based upon the equations of 
rotational hydrodynamics, which provide a description of the condensate in
the Thomas-Fermi (large $g$) regime
\begin{equation}
 \frac{\partial n}{\partial t} + \nabla \cdot (n\bm{v})=0,
\label{eq:hy-dens}
\end{equation}
\begin{equation} 
 \frac{\partial \bm{v}}{\partial t}+(\bm{v}\cdot\nabla)\bm{v} + \nabla
 (V+gn) = 0.
\label{eq:hy-phas}
\end{equation} 
Following \cite{cozzini03} we solve Eqs.\ (\ref{eq:hy-dens}) and 
(\ref{eq:hy-phas}) to describe the dynamics by employing the following ansatz 
for the density and velocity \cite{footnote2}
\begin{equation}
 n (\bm{r})=a_0+a_x x^2 +a_y y^2 + a_{xy} xy,
\label{eq:ans-dens}
\end{equation}
\begin{equation}
 \bm{v} ~(\bm{r})= \bm{\Omega_0} \times \bm{r} + \nabla 
 (b_x x^2+b_y y^2 + b_{xy} xy),
\label{eq:ans-vel}
\end{equation} 
where $a_i$, $b_i$ and $\bm{\Omega_0}$ are time-dependent parameters. Note 
that the velocity (\ref{eq:ans-vel}) includes both rotational and irrotational
components, where the former assumes that the combined velocity field of the
vortices in the lattice approximates a solid body rotation. Substituting 
(\ref{eq:ans-dens}) and (\ref{eq:ans-vel}) into (\ref{eq:hy-dens}) and
(\ref{eq:hy-phas}) yields a set of differential equations for each parameter, 
which are integrated in time using a fourth-order Runge-Kutta scheme
\cite{nr}
\begin{eqnarray}
\label{eq:en1}
 \dot{a}_0 + 2a_0 (b_x+b_y) = 0, \\
\label{eq:en2}
 \dot{a}_x +\Omega_0 a_{xy} + 6a_x b_x + 2 a_x b_x + a_{xy} b_{xy}=0, \\
\label{eq:en3}
 \dot{a}_y-\Omega_0 a_{xy}+2a_y b_x + 6 a_y b_y + a_{xy} b_{xy}=0, \\
\label{eq:en4}
 \dot{a}_{xy}-2\Omega_0 a_x + 2\Omega_0 a_y + 
  4a_{xy} (b_x+b_y) + 2(a_x+a_y) b_{xy} = 0, \\
\label{eq:en5}
 \dot{b}_x+\frac{1}{2} (4b_x^2 - \Omega_0^2 +b_{xy}^2 +1+\epsilon_x+
   2ga_x) = 0, \\	
\label{eq:en6}
 \dot{b}_y+\frac{1}{2} (4b_y^2 - \Omega_0^2 +b_{xy}^2 +1-\epsilon_x+
   2ga_y) = 0, \\
\label{eq:en7}
 \dot{b}_{xy}+2(b_x+b_y)b_{xy} + \epsilon_y +g a_{xy} = 0, \\
\label{eq:en8}
 \dot{\Omega}_0 +2(b_x+b_y) \Omega_0 = 0,
\end{eqnarray}
where $\epsilon_x = \epsilon \cos (2\Omega t)$ and $\epsilon_y =\epsilon \sin 
(2 \Omega t)$, with $\Omega$ the trap rotation frequency. Once the time 
evolution of these properties are known quantities such as the angular momentum
can be calculated by integration of (\ref{eq:ans-dens}) and 
(\ref{eq:ans-vel}).  

\begin{figure}[h]
\centering \scalebox{0.45}
 {\includegraphics{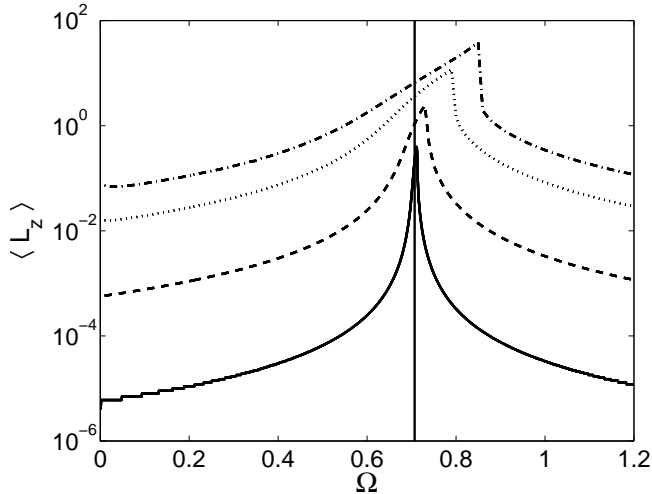}}
 \caption{Maximum angular momentum $\langle L_z \rangle$ as a function of 
 trap rotation frequency, $\Omega$, solving Eqs.\ 
 (\ref{eq:en1}-\ref{eq:en8}) for $\Omega_0=0$. The different curves show
 results for $\epsilon=0.001$ (solid), $\epsilon=0.01$ (dashed), 
 $\epsilon=0.05$ (dotted), and $\epsilon=0.1$ (dot-dashed). The vertical
 line marks $\Omega=1/\sqrt{2}$.} 
\label{fig:peaks}
\end{figure}

The solution of Eqs.\ (\ref{eq:en1}-\ref{eq:en8}) for a rotating elliptical
trap yields an angular momentum
that oscillates in time, similar to the behavior shown in curve b of 
Fig.\ \ref{fig:ang-rot2}. The peak angular momenta as a function of 
$\Omega$ are plotted in Fig.~\ref{fig:peaks} for various values of $\epsilon$
and $\Omega_0=0$. For $\epsilon=0.001$ the response is symmetrical about
a peak near $\Omega=1/\sqrt{2}$. This sharp resonance arises since a 
perturbation rotating with frequency $\Omega$ is resonant with a surface mode 
with azimuthal quantum number $m$ when $\omega - m \Omega \simeq 0$, so that 
for a 
quadrupolar trap deformation the $m=+2$ mode should be resonantly excited when
$\Omega_{\rm res} \simeq \omega_{+2} /2$ \cite{dalfovo00}. With increasing
$\epsilon$ the resonance becomes higher and wider as well as more 
asymmetric, until at $\epsilon=0.1$ there is a steep downward gradient 
immediately following the peak. Note that the resonance peaks are
symmetric when Eqs.\ (\ref{eq:en1}-\ref{eq:en8}) are linearized for 
small departures from equilibrium, demonstrating
that their asymmetry is a consequence of the nonlinearity of the equations.

The results of the hydrodynamic model are compared to those of the GP equation
in Fig.~\ref{fig:angmax}. The solid line plots the 
maximum angular momentum as a function of $\Omega$ for $\Omega_0=0$, where 
one sees a close agreement
with the GP results. Such correspondence is pleasing, but perhaps to be
anticipated when vortices are not nucleated, since Eqs.\
(\ref{eq:ans-dens}) and (\ref{eq:ans-vel}) are expected to be good 
approximations 
and the time dependence of $\langle L_z \rangle$ is oscillatory in both cases.
The agreement is more surprising in the region $0.68 < \Omega < 0.88$, when 
vortices
are nucleated and the time dependence departs from oscillatory behavior. It
appears that the final, steady angular momentum attained is still close to 
the peak amplitude of the oscillations in the absence of vortices.

The maximum $\langle L_z \rangle$ for $\Omega_0=0.5$ is plotted with the
dashed line in Fig.~\ref{fig:angmax}. The agreement with the GP results
is again relatively good, even though the approximation of solid-body rotation
in (\ref{eq:ans-vel}) is not expected to be particularly accurate for 
the small number of vortices present in this case.
One sees a shifting of the resonance to higher trap
rotation frequencies for the larger $\Omega_0$, which is related to 
changes in the frequencies of the quadrupole collective modes. By linearizing
(\ref{eq:en1}-\ref{eq:en8}) for small amplitude oscillations around 
$\epsilon=0$ one finds these frequencies to be \cite{cozzini03}
\begin{equation}
 \omega_{\pm 2} = \sqrt{2-\Omega_0^2} \pm \Omega_0 .
\label{eq:quadfreq}
\end{equation}
This expression provides a simple and useful
way to understand the behavior of quadrupole mode frequencies with changing 
rotation. In particular, one sees that the counter-propagating $m=+2$ and 
$m=-2$ quadrupole modes are degenerate at $\omega=\sqrt{2}$ for $\Omega_0=0$, 
while in the presence of a 
vortex lattice this degeneracy is broken such that $\omega_{+2}\rightarrow 2$ 
and $\omega_{-2} \rightarrow 0$ for $\Omega_0 \rightarrow 1$. This general 
behavior of the $m=+2$ frequency increasing with $\Omega_0$ is reflected in
Fig.~\ref{fig:angmax}, albeit modified to account for the nonlinearity 
introduced with $\epsilon \neq 0$.

\begin{figure}
\centering \scalebox{0.45}
 {\includegraphics{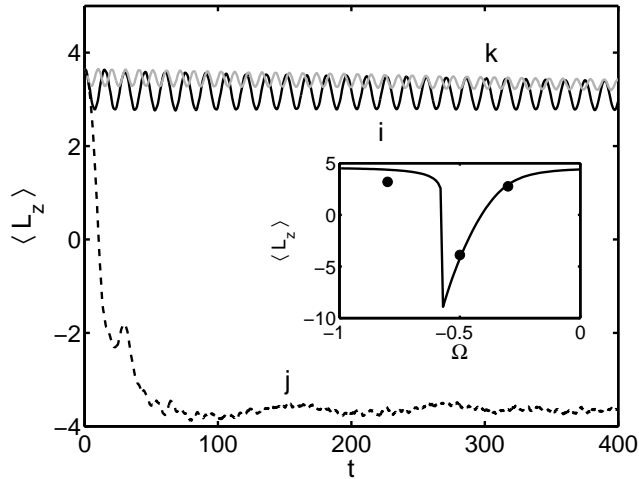}}
 \caption{Angular momentum as a function of time for $\Omega_0=0.5$, and 
 $\Omega=-0.3$ (curve i, solid black line), $\Omega=-0.5$ (j, dashed), and 
 $\Omega=-0.8$ (k, grey). {\it Inset:} Minimum angular momentum for different
 $\Omega$ from GP simulations (points), compared to (solid line) the results 
 of solving Eqs.~(\ref{eq:en1}-\ref{eq:en8}) for $\Omega_0=0.5$.}
\label{fig:angmom3b}
\end{figure}

A further consequence of this argument is that to remove angular momentum,
and therefore vortices, from the condensate requires the trap to be rotating
in the opposite direction with an angular velocity approximately half that
of the $m=-2$ mode frequency. We have confirmed this by performing  
simulations with $\Omega_0=0.5$ and $\Omega \leq 0$, and the results are 
presented in Fig.~\ref{fig:angmom3b}. For $\Omega=-0.3$ the 
angular momentum undergoes small oscillations and remains close to 
the original value $\langle L_z \rangle \simeq 3.64$, since this trap 
rotation rate is sufficiently far from the resonance. This is also found to 
be true when 
$\Omega=-0.8$, albeit with smaller amplitude oscillations due to being further
from resonance. However, when $\Omega=-0.5$ the angular
momentum rapidly decreases and becomes negative as the vortices that are 
originally present
leave the condensate and vortices of opposite sign enter, eventually
settling into a quasi-equilibrium state. The inset of 
Fig.~\ref{fig:angmom3b} compares the minimum of $\langle L_z \rangle$ in 
the three simulations to the result of solving the hydrodynamic model 
(\ref{eq:en1}-\ref{eq:en8}) under the same conditions, once again 
demonstrating reasonable agreement between the two approaches. As expected, 
the peak of 
the resonance is close to $-\omega_{-2}/2$, where $\omega_{-2}\simeq 0.82$
from (\ref{eq:quadfreq}), although as before it is skewed towards higher 
$|\Omega|$ due to the large value of $\epsilon$.

\section{Hysteresis during Linear Ramping}

Finally, we again use numerical simulations of the GP equation 
(\ref{eq:GP-eqn}) to study the difference in the response of the condensate 
between when it is spun up to when it is spun down. We begin with an initial
condition containing vortices, such that the wavefunction is a stationary
solution of Eq.~(\ref{eq:GP-rot}) with $\Omega_0=0.5$. An 
elliptically-deformed trap 
potential ($\epsilon=0.1$) is then rotated with a time-dependent angular 
velocity of the following form
\begin{equation}
 \Omega (t) = \left\{ \begin{array}{ll}
             0.5+3.75 \times 10^{-4}\, t & \ \mbox{if $0\leq t  \leq 400$};\\
             0.8-3.75 \times 10^{-4}\, t & \ \mbox{if $400 < t \leq 800$}.
 \end{array} \right.
\label{eq:rothist}
\end{equation}
Hence during the first half of the simulation $\Omega$ is linearly ramped up 
from 0.5 to 0.65, while in the second half 
it is ramped down from 0.65 to 0.5. These correspond to ``spin-up'' and 
``spin-down'' experiments respectively. 

\begin{figure}
\centering \scalebox{0.48}
 {\includegraphics{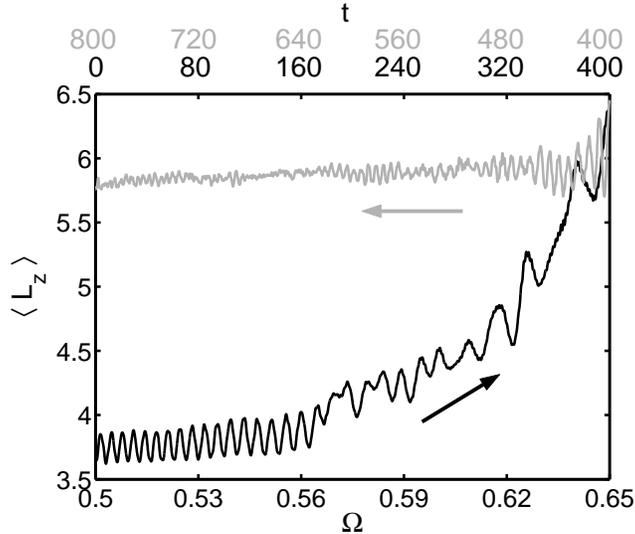}}
 \caption{Hysteresis in the angular momentum, $\langle L_z \rangle$, of the 
 condensate wavefunction, initially a stationary solution of the GP equation 
 (\ref{eq:GP-rot})
 with $\Omega_0=0.5$. The angular velocity of a rotating elliptical trap 
 ($\epsilon=0.1$) follows Eq.~(\ref{eq:rothist}). It is linearly ramped from
 $\Omega=0.5$ at $t=0$ to $\Omega=0.65$ at $t=400$ (black line), and 
 subsequently for $t>400$, is linearly ramped down such that 
 $\Omega=0.5$ at $t=800$ (grey line). Note that the abscissa is labelled with 
 $\Omega$ (bottom) and $t$ (top), where the latter has two different 
 tick labels for spin-up (black) and spin-down (grey).}
\label{fig:hyst}
\end{figure}

The result is shown in Fig.~\ref{fig:hyst}, where the angular momentum is 
plotted against both $\Omega$ (lower abscissa) and time (upper abscissa). In 
turn, on the upper abscissa times are shown for both the spin-up (black text) 
and spin-down (grey text) 
processes, where the corresponding plotted curves are represented by the 
same colors. During the spin-up 
process $\langle L_z \rangle$ does not increase initially, but undergoes 
oscillatory behavior. This corresponds to the left side of 
Fig.~\ref{fig:angmax}, where the rotation frequency is far from the $m=+2$
quadrupolar resonance. Then, as $\Omega$ approaches resonance angular momentum
is transferred to the condensate, until $\langle L_z \rangle \simeq 6$ at 
$t=400$. During the 
subsequent spin-down process, however, the angular momentum does not follow
the path of the spin-up, but remains almost constant with only a 
slight decrease at the end. This reflects the fact that the rotation rate 
is far from being resonant with the $m=-2$ mode, the excitation of which is 
required to remove angular momentum from the system.
Once again this clearly illustrates the importance of 
the rotation history on the attained angular momentum, and hence 
demonstrates the possibility of observing hysteresis phenomena in rotating 
condensates.    

Finally, we should note that hysteresis behavior was also found by 
Garc\'{i}a-Ripoll and P\'{e}rez-Garc\'{i}a in Ref.\ \cite{garciaripoll01a}.
A major difference with respect to the present work is that in 
\cite{garciaripoll01a} the time-independent GP equation was solved to find
the stationary solutions for different rotation rates, while here the 
dynamics are considered explicitly by solving the time-dependent GP equation.
Moreover, the results of Ref.\ \cite{garciaripoll01a} are based on the 
assumption that all collective modes are excited, and vortex formation occurs
when the rotation frequency $\Omega$ first becomes resonant with one of the 
modes. In experiments
this would correspond to relatively high temperatures where the collective
excitations are thermally populated to an appreciable extent and the thermal 
cloud is in rotation (similarly to the experiment of Haljan {\it et al.} 
\cite{haljan01}). In contrast, we have been interested in the case where vortex
nucleation is induced by a rotating elliptical trap, so that the quadrupole
mode is predominately excited and becomes unstable prior to vortex nucleation.
  
\section{Conclusions}

In summary, we have studied the process of vortex formation in a rotating 
dilute Bose-Einstein condensate using numerical simulations of the 
Gross-Pitaevskii equation. We have paid particular attention to the amount of 
angular momentum transferred to the condensate
for different rotation angular velocities of an elliptically-deformed
trapping potential. The angular momentum of
the condensate (and hence the number of vortices) not only depends upon the 
final rotation rate of the trap, but also upon the history of the rotation. 
Specifically, 
by initially rotating at one angular velocity, $\Omega$, until vortices are 
nucleated, then changing to a second value of $\Omega$, one can attain
a very different angular momentum compared to that achieved when the second 
$\Omega$ is retained throughout. Furthermore, we have shown that simulations 
with different initial conditions, corresponding to when vortices are already 
present in the condensate, also lead to different final angular momenta, 
similar to behavior found in the experiment of Ref.\ \cite{bretin04}.  
Using a hydrodynamic model we have  
demonstrated that this effect is due to a shift of the resonant frequency
for the excitation of quadrupole collective modes when the condensate is 
already in rotation. We also show that the 
angular momentum
response to a linear ramp of the trap rotation frequency is radically 
different depending on whether the ramp is increasing or decreasing, revealing
the existence of hysteresis phenomena in rotating condensates.  

We conclude by noting that these processes should be readily 
observable in present experiments where vortices have been nucleated by 
stirring the condensate (e.g.\ Refs.~\cite{madison00,abo-shaeer01,hodby02}), 
although the dissipation arising from the presence of a non-condensed 
cloud may influence the angular momentum eventually 
achieved over longer time scales. Assessing the importance of this 
contribution requires a more 
sophisticated model that includes finite temperature effects 
\cite{zaremba99,jackson02}, and will be left to future work. 

\acknowledgments

Financial support for this research was provided by the Ministero 
dell'Instruzione, dell'Universit\'{a} e della Ricerca, and the Engineering and
Physical Sciences Research Council. We gratefully acknowledge useful 
discussions with C.~Adams, A.~Fetter, N.~Parker and N.~Proukakis.


\begin{thebibliography}{99}
\bibitem{footnote1}
 Note that for $\sqrt{1-\epsilon}<\Omega<\sqrt{1+\epsilon}$ motion of the 
 condensate center-of-mass is dynamically unstable \cite{rosenbusch02}, such
 that its displacement from the center of the trap grows exponentially in 
 time. This precludes simulations in the range $0.95<\Omega<1.05$ for 
 $\Omega_0=0.5$, since the small (but non-zero) initial displacement of 
 the center-of-mass leads to the condensate rapidly spiralling out of the 
 numerical grid. On the other hand, for $\Omega_0=0$ the initial displacement 
 is very 
 close to zero, so that the instability is less of a problem over applicable
 timescales, and simulations are possible in this range.
\bibitem{footnote2}
 Eqs.\ (\ref{eq:ans-dens}) and (\ref{eq:ans-vel}) can be easily extended 
 to three dimensions by including the additional parameters $a_z$ and $b_z$, 
 as in \cite{cozzini03}. Similar results are found, further verifying
 that the 2D approach taken here is reasonable.
\end{thebibliography}
\end{document}